\documentclass[%
 reprint,
showpacs,
 amsmath,amssymb,
 aps,
pra,
]{revtex4-1}

\usepackage{longtable}
\usepackage{graphicx}
\usepackage{dcolumn}
\usepackage{bm}
\usepackage{hyperref}
\usepackage[mathlines]{lineno}

\begin{document}

\preprint{APS/123-QED}

\title{Monte Carlo determination of the critical coupling in \texorpdfstring{$\phi^4_2$}{} theory}

\author{Paolo Bosetti$^1$}
 \email{paolo.bosetti01@universitadipavia.it}
\author{Barbara De Palma$^{1,2}$}%
 \email{barbara.depalma@pv.infn.it}

\author{Marco Guagnelli$^{1,2}$}
 \email{marco.guagnelli@pv.infn.it}

\affiliation{
$^1$ Dipartimento di Fisica, Universit\`a degli Studi di Pavia, Via A. Bassi 6, 27100, Pavia, Italy \\
$^2$ INFN, Sezione di Pavia, Via A. Bassi 6, 27100, Pavia, Italy}%

\date{\today}

\begin{abstract}
We use lattice formulation of $\phi^4$ theory in order to investigate non--perturbative features of its continuum limit in two dimensions. In particular, by means of Monte Carlo calculations, we obtain the critical coupling constant $g/\mu^2$ in the continuum, where $g$ is the {\em unrenormalised} coupling. Our final result is $g/\mu^2=11.15 \pm 0.06_{stat} \pm 0.03_{syst}$.
\end{abstract}

\maketitle


\section*{Introduction}

$\phi^4$ theory plays a phenomenological role as an extremely simplified model for the Higgs sector of the Standard Model. In \cite{ref1:Aizenman,ref2:Frohlich} the triviality of $\phi^4$ theory in more than four dimensions has been proven, and there are numerous analytical and numerical results for $D=4$ \cite{ref3:Luscher,ref4:Brezin,ref5:Wolff}, indicating that in this case the theory is trivial as well.

In $D=2$ and $D=3$ the theory is super--renormalisable: the coupling constant has positive mass dimensions. In this paper we will work in $D=2$, employing lattice regularisation. In $D=2$, $[g]=[\mu_0^2]$, where $\mu_0$ is the (bare) mass parameter of the theory. This means that the only physically relevant dimensionless parameter is the ratio $g/\mu^2$, where $g$ is the bare coupling constant and $\mu^2$ is a renormalised squared mass in some given renormalisation scheme. An additive mass renormalisation is required since in the continuum limit the bare mass parameter diverges like $\log(a)$, where $a$ is the lattice spacing. We do not care about coupling renormalisation, since it amounts to a finite factor.

Despite the simplicity of the model, there is still debate in the literature about the value of $f \equiv g/\mu^2$, where the ratio is evaluated at the critical point. In particular we are interested in the value of $f$, call it $f_0$, computed in the limit in which both $g$ and $\mu^2$ go to zero; this corresponds to the critical value in the continuum. We decided to tackle this problem by using the same renormalisation scheme used in \cite{ref7:LoinazWilley,ref6:SchaichLoinaz}, adopting the simulation technique introduced in \cite{ref19:WolffConf}, namely the \emph{worm algorithm}, and using a completely different strategy to obtain $g/\mu^2$ in the infinite volume limit.

In the following we will describe the model and the renormalisation scheme chosen in order to extract $\mu^2$ at fixed $g$ in the infinite volume limit from our simulations. Then we will give details about the simulations and we will proceed to the continuum limit extrapolation. In the end we will compare our results with recent determinations of the same quantity and we will draw some conclusions. \\

\section{Lattice Formulation}

Let's introduce the $\phi^4$ Lagrangian in the Euclidean space:
\begin{equation}
\mathcal{L}_E = \dfrac{1}{2}\left(\partial_\nu\phi \right)^2 + \dfrac{1}{2}\mu_0^2\phi^2 + \dfrac{g}{4}\phi^4.
\end{equation}
In $D=2$ the Euclidean action is
\begin{equation*}
\mathcal{S}_E = \int d^2x\,\mathcal{L}_E.
\end{equation*}
In order to obtain a dimensionless discretized action we put the system on a 2-dimensional lattice with spacing $a$ and introduce the following parametrization
\begin{equation}
\label{ga-mua}
\hat{\mu}^2_0=a^2\mu^2_0, \qquad \hat{g}=a^2g.
\end{equation}
In this way we have
\begin{equation}
\mathcal{S}_E =  \sum_x \left\lbrace -\sum_{\nu}\phi_x\phi_{x+\hat{\nu}} + \dfrac{1}{2}\left( \hat{\mu}_0^2+4\right)\phi_x^2 + \dfrac{\hat{g}}{4}\phi_x^4 \right\rbrace, 
\end{equation}
where $\phi_{x\pm\hat{\nu}}$ are fields at neighbor sites in the $\pm\nu$ directions. 

In the following we will omit the ``hat'' on top of lattice parameters: all quantities will be expressed in lattice units, {\em i.e.} they become dimensionful when multiplied by appropriate powers of the lattice spacing $a$.

If we take the continuum limit too naively, at fixed physical quantities, we obtain, in $D<4$, the critical Gaussian model \cite{ref9:Parisi}. On the other hand, if we stick to a fixed value of $g$ (in lattice units) we can search for a value of $\mu_0^2$ such that we get, in the infinite volume limit, a second order phase transition point in the plane $(g,\mu^2_0)$. 

\begin{figure}[h]
\centering
\includegraphics[width=0.25\textwidth]{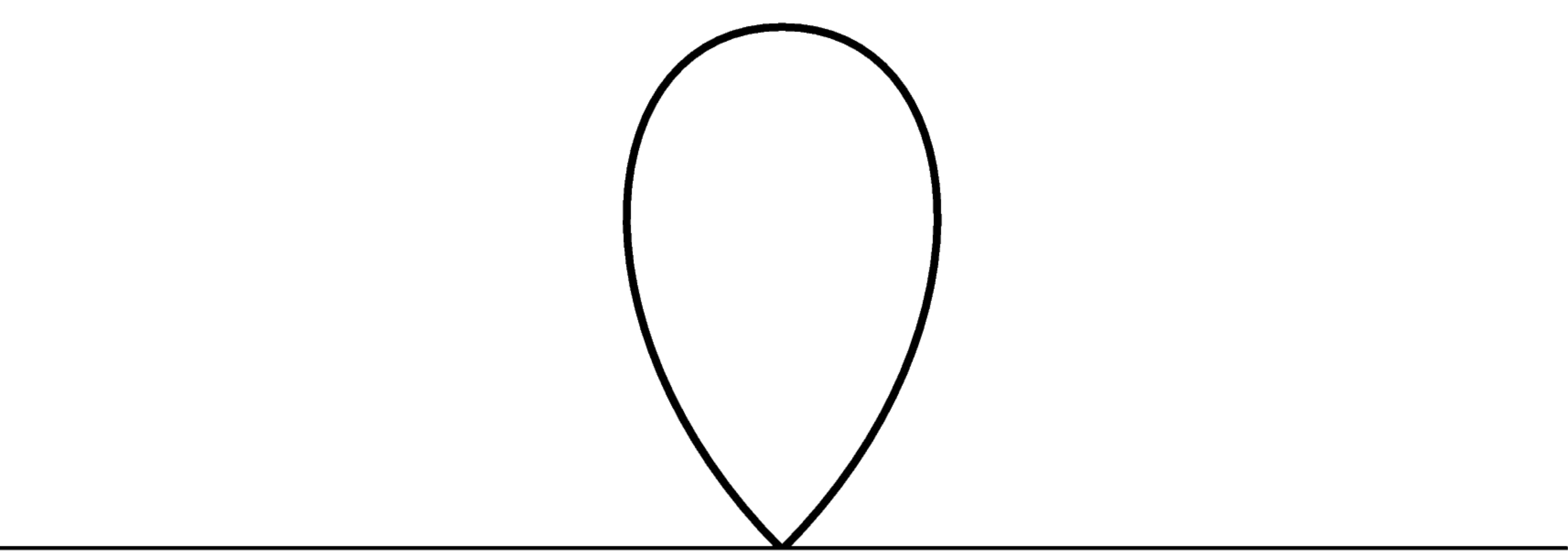}
\caption{\label{fig:tadpole}One--loop self--energy in $\phi^4$}

\end{figure}

In order to safely go to the continuum limit, we have to work out an additive renormalisation of the mass parameter, since $\mu_0^2$ in this limit diverges like $\log(a)$; in this way we translate $\mu^2_0$ into $\mu^2$, a renormalised squared mass. Of course several definitions of renormalised mass can be chosen; in this work we adhere to the same renormalisation procedure as in \cite{ref7:LoinazWilley,ref6:SchaichLoinaz}. We refer the reader to these papers for more details. Here we only remind that in $D=2$ there is only a 1--Particle--Irreducible divergent diagram (see Fig. \ref{fig:tadpole}). Its expression on a lattice with $N\times N$ points is
\begin{equation}
A(\mu^2_0) = \dfrac{1}{N^2}\sum_{k_1=0}^{N-1}\sum_{k_2=0}^{N-1}\dfrac{1}{4\left(\sin^2 \dfrac{\pi k_1}{N} + \sin^2 \dfrac{\pi k_2}{N}\right) + \mu^2_0},
\end{equation}
and a suitable renormalisation condition consists in putting $\mu^2$ equal to the solution, in the infinite volume limit, of the equation
\begin{equation}
\mu^2 = \mu^2_0 + 3gA(\mu^2).
\label{eq:defmu2}
\end{equation}

This condition is equivalent to the introduction of a proper divergent mass--squared counterterm in the action. 
We may finally extrapolate the quantity $f\equiv g/\mu^2$ to $g\to 0$ in order to obtain $f_0$, the critical value in the continuum limit. 

Another parametrization of the action is the following:
\begin{equation}\label{eq:sbetalambda}
\begin{split}
\mathcal{S}_E &= -\beta\sum_x\sum_\nu\varphi_x\varphi_{x+\hat{\nu}} + \sum_x\left[\varphi^2_x + \lambda(\varphi^2_x-1)^2\right] \\
&= \mathcal{S}_{I} + \mathcal{S}_{Site},
\end{split}
\end{equation}
where the relations between $(\mu_0^2,\;g)$ and $(\beta,\;\lambda)$ are:
\begin{equation}\label{param}
\phi_x = \sqrt{\beta}\varphi, \qquad \mu_0^2 = 2\dfrac{1-2\lambda}{\beta}-4, \qquad g=\dfrac{4\lambda}{\beta^2}.
\end{equation} 

In eq.(\ref{eq:sbetalambda}) there is an interaction term between neighbor sites, $\mathcal{S}_{I}$, with a coupling constant of strength $\beta$ and a term related to a single site, $\mathcal{S}_{Site}$. With this parametrization it is easy to recognize the \emph{Ising limit} for $\lambda\to\infty$. In this limit, configurations with $\varphi^2 \neq 1$ are completely suppressed and the fields assume only values $\varphi(x)=\pm 1$. As a result, the second term of \eqref{eq:sbetalambda} can be disregarded and the action becomes the well-known Ising action $\mathcal{S}_E = -\beta\sum_x\sum_\nu\varphi_x\varphi_{x+\hat{\nu}}$.

\subsection{Simulations}

In this section we outline our general computational strategy, postponing the discussion of the simulations details.

We use the \emph{worm algorithm} \cite{ref19:WolffConf}, using the lattice action given by (\ref{eq:sbetalambda}). We checked our simulation program against the results of \cite{ref19:WolffConf}\footnote{We refer the reader to this paper for all details of the algorithm itself.}, obtaining values compatible within errors, well below one sigma level. In this case and also in the following, in order to estimate statistical errors we use the program described in \cite{ref14:lessErr}.

Considering a fixed value of $\lambda$, our aim is to compute the critical point of the theory, {\em i.e.} the critical value of $\beta$ for that particular value of $\lambda$. We use the physical condition
\begin{equation}\label{mL}
mL = L/\xi = \textrm{const} = z,
\end{equation}
where $m$ is implicitly defined by the condition
\begin{equation}\label{eq:defm}
\dfrac{G(p^*)}{G(0)} = \dfrac{m^2}{p^{*2} + m^2}. 
\end{equation}
$G(p)$ is the two--point function in momentum space, and $p^*$ is the smallest possible momentum on a lattice of linear size $L$. Details, as before, in \cite{ref19:WolffConf}. Condition \eqref{mL} implies that $\xi$ grows linearly with $L$, and when $L/a \to\infty$ we arrive at the critical point.
We then simulate several lattices with different values of $N \equiv L/a$; for each couple $(\lambda,\,N)$ we obtain a value of $\beta(\lambda,\,N)$ such that $mL = z$. After this step we extrapolate our results to $a/L\to 0$ in order to compute $\beta(\lambda)$. Now, using relations in \eqref{param} we derive $g(\lambda,\beta)$ and $\mu_{0}^2(\lambda,\beta)$. Using renormalisation condition \eqref{eq:defmu2} we finally pin down $\mu^2(g)$ and hence the ratio $f \equiv g/\mu^2$. 

We repeat all this procedure for several values of $\lambda$, and hence of $g$; in the end we extrapolate our results to $g\to 0$, in order to obtain $f_0$. We will now focus on the details of our simulations. \\

We choose the condition $z = 4$. As we will see in the following, this choice is not as crucial as it may seem.

\begin{figure}[h]
\begin{center}
\includegraphics[width=0.50\textwidth]{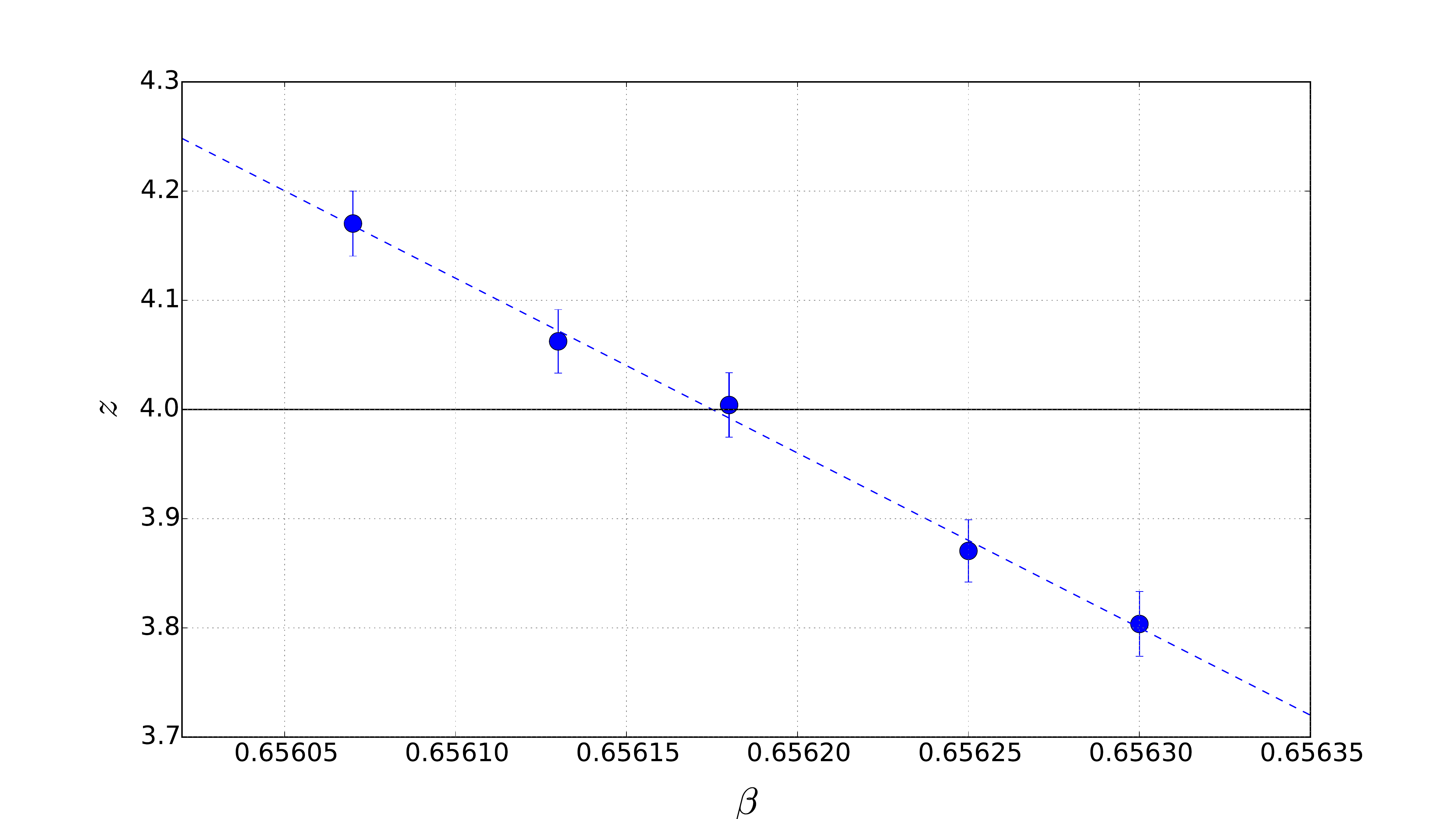}
\caption{\label{fig:beta}Linear interpolation ($\lambda = 0.25,\,L=256$) to obtain $\beta(z=4)$}

\end{center}
\end{figure}

At a fixed value of $\lambda$ we simulate the system for five values of $L/a$, namely: $L/a=$ 192, 256, 384, 512 and 768. For each value of $L/a$ few preliminary simulations are needed to roughly find the value of $\beta$ leading to $z \simeq 4$. In few cases (see for example Fig. \ref{fig:beta}) we have explicitly checked that using five values of $\beta$ such that $z$ falls approximately into the interval $[3.8, 4.2]$ we do not observe any sign of non--linearity of $z$ as a function of $\beta$. The difference in $\beta(z=4)$ between the case in which we use $5$ points to interpolate and the case in which we use only $3$ points is one order of magnitude less than the statistical error itself. We then decided to use just $3$ values of $\beta$ for the real simulations to linearly interpolate the results and to obtain in this way $\beta(\lambda,N)$. 

A typical full simulation ($\lambda = 0.25$) is synthesized in Table \ref{full025}.
\begin{table}[h]
\caption{\label{full025}$\lambda=0.25$ simulations}

\begin{ruledtabular}
\begin{tabular}{cccc}
$L/a$    & $N_{meas}$     &  $N_{sweep}$ & $\beta_c(z=4)$    \\
\hline
$192$    & $1\times 10^5$ &  $15$        &  $0.655357(12)$ \\
$256$    & $5\times 10^4$ &  $15$        &  $0.656177(11)$ \\
$384$    & $5\times 10^4$ &  $15$        &  $0.656984(8)$ \\
$512$    & $3\times 10^4$ &  $20$        &  $0.657399(7)$ \\
$768$    & $2\times 10^4$ &  $25$        &  $0.657818(10)$ \\
\hline
$\infty$ &                &              &  $0.658628(10)$ \\
\end{tabular} 
\end{ruledtabular}
\end{table}
\noindent
$N_{sweep}$ is the number of worm--sweeps between two measures, which increases in order to minimize the simulation time, taking into account autocorrelation time; the number of thermalisation sweeps for all our simulations is several hundreds times $\tau$, the autocorrelation time of $mL$, which we always keep under control.\\

\begin{figure}[h]
\begin{center}
\includegraphics[width=0.5\textwidth]{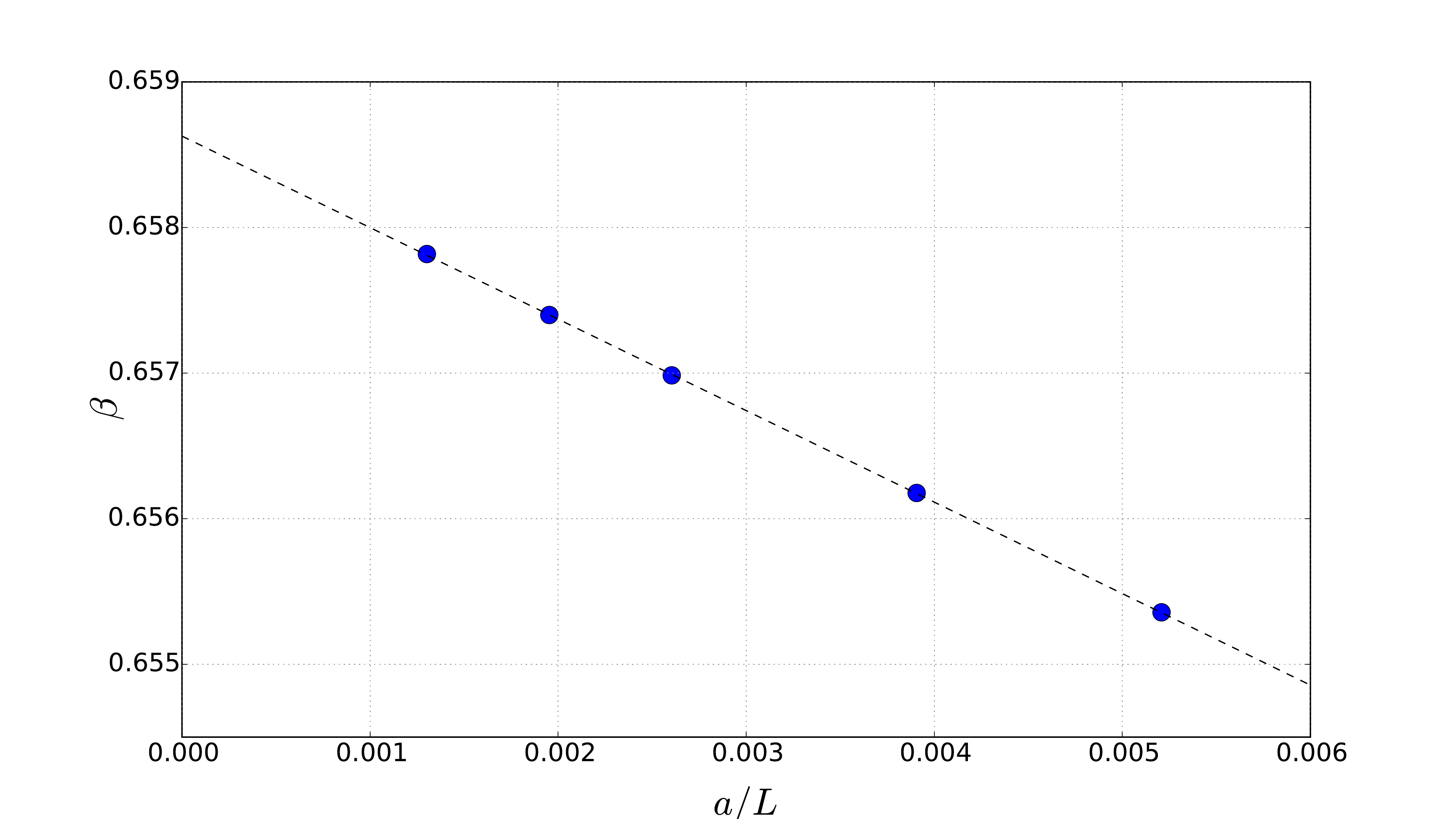}
\caption{\label{fig:betac}Linear extrapolation of $\beta$ to $a/L=0$ for $\lambda=0.25$.}

\end{center}
\end{figure}

$\phi^4$ theory \cite{ref11:universality} is in the same universality class of the Ising model, and we know that in $D=2$ the critical exponent of the correlation length is $\nu=1$. Thanks to finite size scaling arguments we expect to be able to extrapolate $\beta(\lambda, N)$ to $\beta(\lambda)$ linearly in $a/L$. This is numerically very well confirmed for all values of $\lambda$ we explored. In Fig. \ref{fig:betac} we show a typical extrapolation. For every value of $\lambda$ considered, we obtain a very reasonable value of $\chi^2 \le 1$.
Our final results are reported in Table \ref{table:FinalRes}.\\ 
\begin{table*}[ht!]
\caption{\label{table:FinalRes}Final extrapolations to infinite volume limit: $g$ and $\mu^2$ are computed at $\beta_c$ using equations \eqref{eq:defmu2} and \eqref{param}.}

\begin{ruledtabular}
\begin{tabular}{ccccc}
$\lambda$ & $\beta_c$ & $g$ & $\mu^2$ & $g/\mu^2$  \\
\hline
$1.000000$ &  $0.680601(11)$ & $8.63523(29)$ & $0.649451(67)$ & $13.2962(18)$  \\
$0.750000$ &  $0.689117(13)$ & $6.31733(24)$ & $0.509730(59)$ & $12.3935(19)$ \\
$0.500000$ &  $0.686938(10)$ & $4.23833(12)$ & $0.367173(31)$ & $11.5431(13)$  \\
$0.380000$ &  $0.678405(11)$ & $3.30267(10)$ & $0.296195(32)$ & $11.1503(15)$  \\
$0.250000$ &  $0.6586276(98)$ & $2.305261(69)$ & $0.214762(27)$ & $10.7340(17)$  \\
$0.200000$ &  $0.6462478(78)$ & $1.915543(46)$ & $0.181077(21)$ & $10.5786(15)$  \\
$0.125000$ &  $0.6190716(52)$ & $1.304633(25)$ & $0.125924(15)$ & $10.3605(15)$  \\
$0.094000$ &  $0.6030936(89)$ & $1.033757(30)$ & $0.100518(23)$ & $10.2843(26)$  \\
$0.062500$ &  $0.5820989(60)$ & $0.737813(15)$ & $0.072073(15)$ & $10.2370(23)$  \\
$0.030000$ &  $0.5516594(71)$ & $0.394311(10)$ & $0.038407(17)$ & $10.2666(48)$  \\
$0.015625$ &  $0.5326936(27)$ & $0.2202547(22)$ & $0.0211916(63)$ & $10.3935(32)$  \\
$0.007500$ &  $0.5187729(29)$ & $0.1114722(12)$ & $0.0105457(67)$ & $10.5704(68)$  \\
$0.005000$ &  $0.5136251(17)$ & $0.07581192(49)$ & $0.0071014(38)$ & $10.6757(57)$  \\
$0.002000$ &  $0.5064230(16)$ & $0.03119343(19)$ & $0.0028637(35)$ & $10.8925(132)$  \\
\end{tabular} 
\end{ruledtabular}
\end{table*}
 
\begin{figure}[h]
\begin{center}
\includegraphics[width=0.5\textwidth]{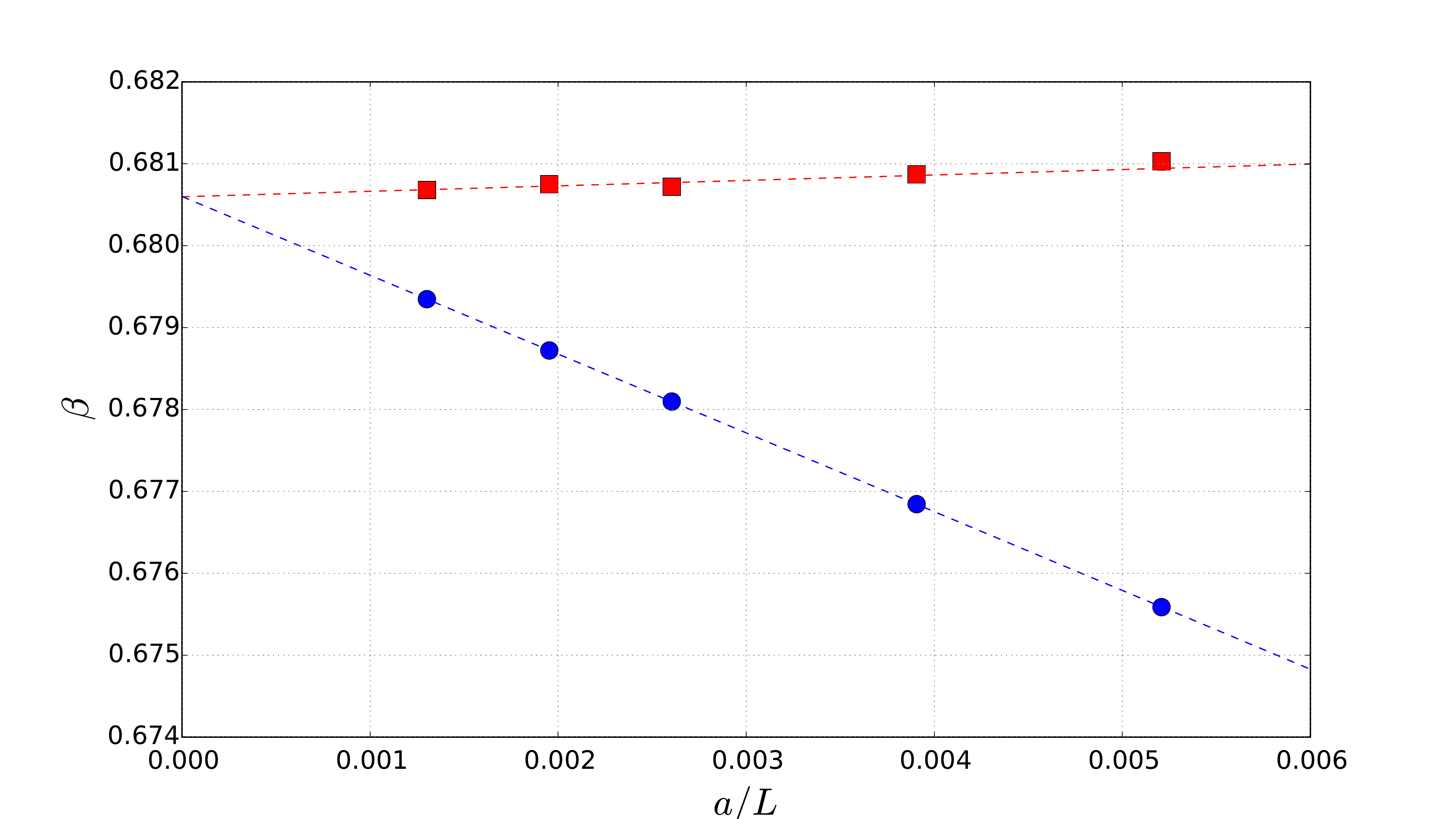}
\caption{\label{fig:z1z4}Extrapolation to $a/L=0$ with $z=4$ (blue steep curve)  and $z=1$ (red curve) ($\lambda=1$). }
\end{center}
\end{figure}

Now we show that the condition $z=4$ is not crucial; actually, as is well known from general theoretical arguments, we could choose another value of $z$ without affecting the results in the infinite volume limit. From a numerical point of view it is nevertheless interesting to consider other values of $z$ in order to be more confident on the reliability of the extrapolations. As an example we show, in Fig. \ref{fig:z1z4}, a double extrapolation to $a/L = 0$ in the case $\lambda=1$. For $z=4$ the extrapolation to $a/L=0$ is steeper than for $z=1$, since in the latter case, at finite volume, we are nearer to criticality, so that $\beta(\lambda,N)$ is not so far from the infinite volume value. Nevertheless at $z=4$ we obtain a much more clear signal; we can extrapolate to the $a/L=0$ value with a much smaller statistical error even if the number of measures is $(5-10)$--times smaller than the case $z=1$. The results in the infinite volume limit coincide within the statistical errors; $\beta(z=1) = 0.68060(4)$, to be compared with the equivalent value in Table \ref{table:FinalRes}, $\beta(z=4) = 0.680601(11)$.

\section{Results}

\begin{figure}[h]
\begin{center}
\includegraphics[width=0.5\textwidth]{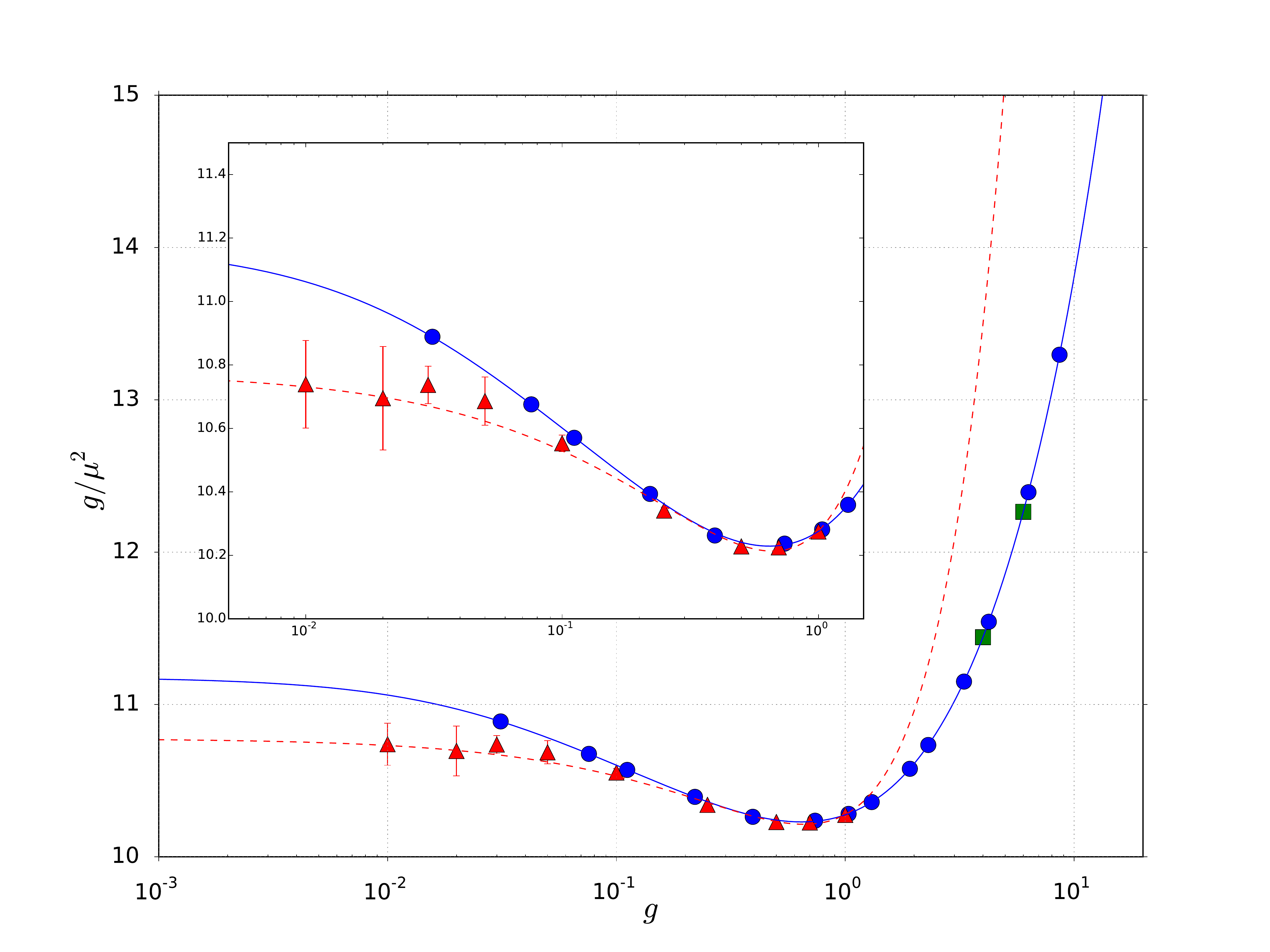}
\caption{\label{fig:tot2}Final results for $f(g)$ in logarithmic scale. Error-bars, where not visible, are smaller than symbols size.}

\end{center}
\end{figure}

In Fig. \ref{fig:tot2} we plot the results shown in Table \ref{table:FinalRes}. The plot is in $x$--log scale, to emphasize the fact that we covered over two order of magnitude in $g$. Blue round points are our results taken from Table \ref{table:FinalRes}. Red triangular points are results from \cite{ref6:SchaichLoinaz}. We postpone the discussion of the green square points.


First of all we note that in the intermediate region, {\em i.e.} in the minimum of the curve, our results are in almost perfect agreement with those of \cite{ref6:SchaichLoinaz}. Note that the infinite volume limit results of \cite{ref6:SchaichLoinaz} are obtained with a completely different strategy. The situation starts changing at the lowest simulated values of $g$: we see, in the insert shown in Fig. \ref{fig:tot2}, that our points seem to be a little bit higher. The blue curve is our final fitting function, which we are now going to discuss, while the red dashed curve is the fit function used in \cite{ref6:SchaichLoinaz}.

\begin{figure}[h]
\begin{center}
\includegraphics[width=0.5\textwidth]{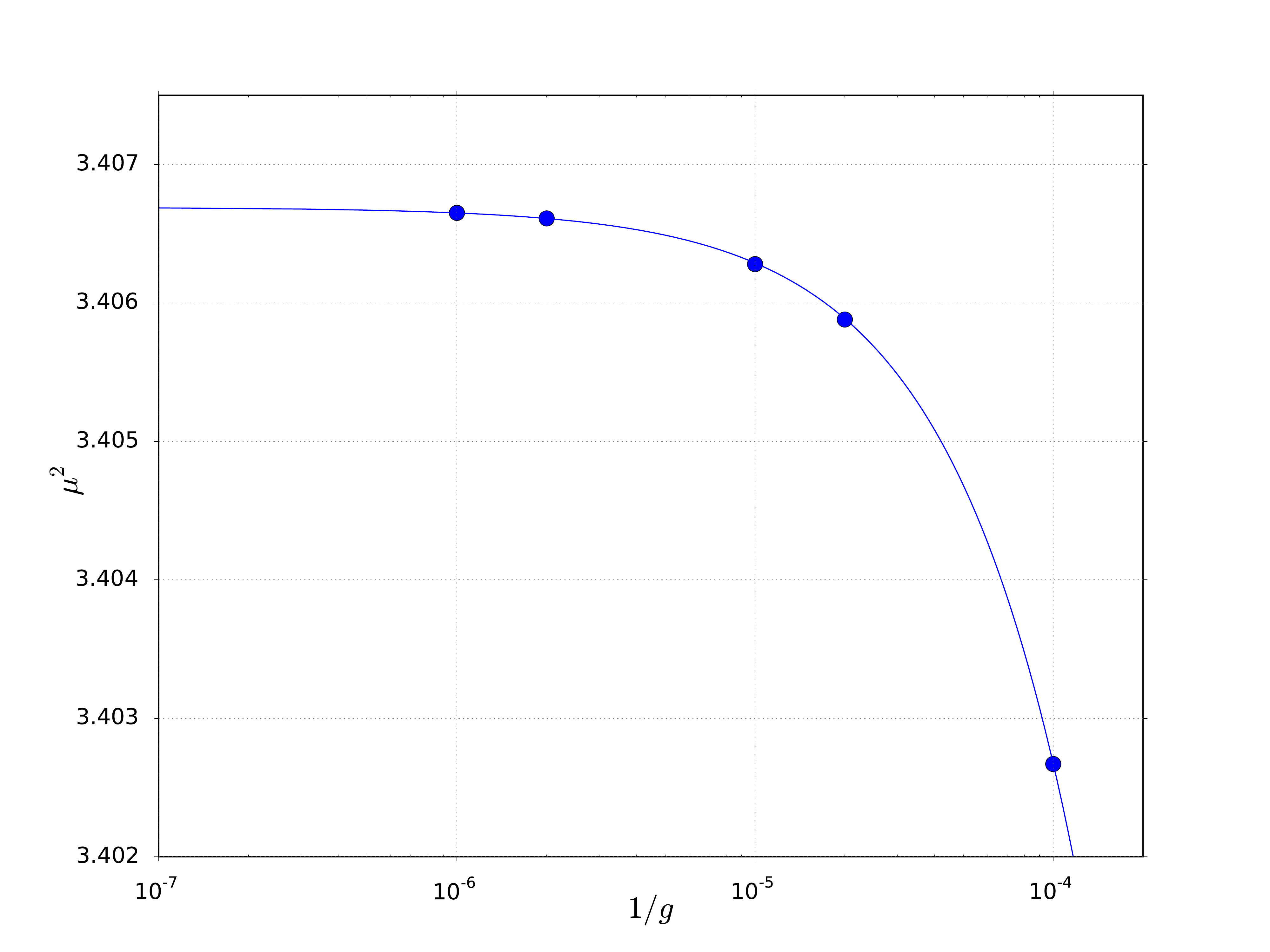}
\caption{\label{fig:ising}Extrapolation of $\mu^2$ at $g\to\infty$, as described in text.}

\end{center}
\end{figure}

We decided to fit $f(g)$ over the entire range at our disposal with the function
\begin{equation}
\label{eq:fitfg}
f(g) = \dfrac{a_0 + a_1g + a_2g^2 + a_3g^3 + a_4g^4}{1 + b_1g + b_2g^2 + b_3g^3}.
\end{equation}
We can certainly justify the functional form for large values of $g$. We know that $\phi^4$ theory reduces to the Ising model in the limit $\lambda\to\infty$. In particular in the Ising limit we have $\beta = \beta_c^{\textrm\footnotesize{Ising}} = \dfrac{\log(1+\sqrt{2})}{2}$. Note that $\beta(\lambda)$, at the critical point, is a highly non--linear function of $\lambda$ itself. In fact at $\lambda=0$, $\beta=0.5$; then we note a maximum, with a value around $0.69$ for intermediate values of $\lambda$; in the end $\beta(\lambda)$ has to go asymptotically to the value $0.44068679\dots$, the critical Ising value in $D=2$. In \cite{ref12:scaling} it is noted that for $\lambda=10$ the value of $\beta$ at criticality is already near the asymptotic value. For very large values of $\lambda$ we can then safely approximate $\beta$ with $\beta_c^{\textrm\footnotesize{Ising}}$; if we look at the relations \eqref{param}, we note that $g$ is going to infinite linearly with $\lambda$, and $\mu_0^2$ diverges proportionally to $g$. But this is not true for $\mu^2$ due to the renormalisation condition \eqref{eq:defmu2}. We numerically checked that $\mu^2$, using the approximation $\beta = \beta_c^{\textrm\footnotesize{Ising}}$ for $g \ge 10^4$, can be linearly extrapolated in $1/g$ to $g\to\infty$ (see Fig. \ref{fig:ising}). We arrive at the value $\mu^2_{\textrm\footnotesize{Ising}} = 3.40669(1)$; the error is subjectively estimated from the fit.

We simply assume a linear behavior of $f(g)$ for $g\to 0$. Taking into account the Ising limit constraint, we fix the parameter $b_3$ as a constant times $a_4$. We have in total $7$ d.o.f. and we obtain 
\begin{equation}
f_0 = 11.179(62)
\label{eq:f1}
\end{equation}
with a reduced $\chi^2 = 0.73$. 

In order to check the validity of the fit function \eqref{eq:fitfg}, we decided to compute $f(g)$ with the same strategy adopted in \cite{ref6:SchaichLoinaz}, but for two values of $g$ higher than those considered in \cite{ref6:SchaichLoinaz}, namely $g=4$ and $g=6$. The field configurations are generated with a mixture of Metropolis steps and single cluster Wolff steps, used in \cite{ref6:SchaichLoinaz} and presented in \cite{ref13:WClusterAlg}.

In particular for each $L/a$ we search for the value of $\mu^2_0$ that maximize the magnetic susceptibility $\chi = \langle \bar\phi^2 \rangle - \langle |\bar\phi| \rangle^2$; this peak is a signal of the pseudo--transition point at finite volume. $\bar\phi$ is the average of the field over the whole lattice. $\mu^2_0$ is then extrapolated to $a/L\to 0$ and the corresponding $\mu^2$ is obtained by means of condition \eqref{eq:defmu2}.

Details of simulations for $g=4$ are given in Table \ref{tab:g4}.
\begin{table}[ht!]
\caption{\label{tab:g4}$g=4$ simulations with Metropolis--cluster algorithm}

\begin{ruledtabular}
\begin{tabular}{cccc}
$L/a$    & $N_{meas}$     &  $g/\mu^2$     \\
\hline
$128$    & $1\times 10^5$ &  $11.2631(13)$        \\
$192$    & $1\times 10^5$ &  $11.3227(9)$         \\
$256$    & $1\times 10^5$ &  $11.3533(7)$         \\
$384$    & $1\times 10^5$ &  $11.3826(3)$         \\
$512$    & $1\times 10^5$ &  $11.3969(3)$         \\
\hline
$\infty$ &                &  $11.4417(5)$         \\
\end{tabular} 
\end{ruledtabular}
\end{table}
\noindent
As can be seen in Fig. \ref{fig:tot2} the two points at $g=4$ and $g=6$, represented by squares, lie perfectly on the curve defined by our fit function. This represent a further confirmation that our strategy for computing $g/\mu^2$, passing through the limiting procedure described above, works as expected. \\
In order to better understand the behavior of $f(g)$ for all possible values of $g$ we define a new parameter, $\eta$:
\begin{equation}
\label{eq:eta}
\eta = \dfrac{g}{g+1}.
\end{equation}
It is clear that \eqref{eq:eta} is a map from $g\in [0,\infty)$ to $\eta\in [0,1]$. We hope in this way to obtain a smoother behavior of $f(\eta)$; note that the limit $f(\eta\to 0)$ is completely equivalent to $f(g\to 0)$. We then define the fit function
\begin{equation}
\label{eq:feta}
f(\eta) = \dfrac{a'_0 + a'_1\eta + a'_2\eta^2 + a'_3\eta^3}{1 + b'_1\eta + b'_2\eta^2 + b'_3\eta^3},
\end{equation}
where one of the parameters is determined by the Ising constraint for $\eta = 1$.

\begin{figure}[h]
\begin{center}
\includegraphics[width=0.5\textwidth]{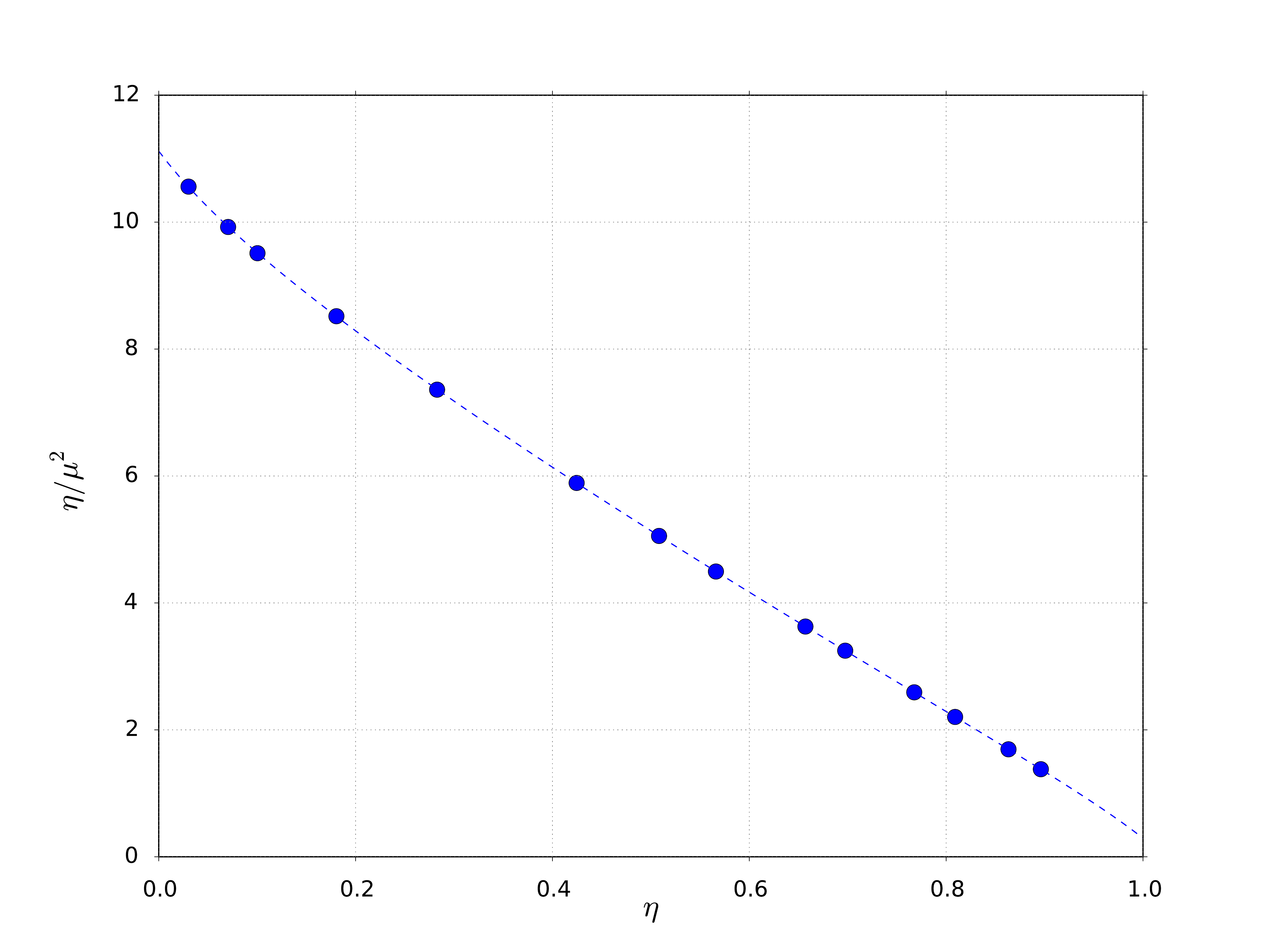}
\caption{\label{fig:feta}Final plot of $f(\eta)$ with our results.}

\end{center}
\end{figure}

As shown in Fig. \ref{fig:feta}, this choice leads us to a smoother function. With the $\eta$ parametrization we obtain:
\begin{equation}
f_0 = 11.119(24),
\label{eq:f2} 
\end{equation}
with a reduced $\chi^2=0.95$ and $8$ d.o.f.

\begin{table}[hb!]
\caption{\label{tab:review}Sample of the results for the continuum critical parameter $f_0$ from the literature. DLCQ stands for \emph{Discretized Light Cone Quantization}, QSE diagonalization for \emph{Quasi--Sparse Eigenvector} diagonalization and DMRG for \emph{Density Matrix Renormalization Group} }

\begin{ruledtabular}
\begin{tabular}{lcc}
Method					& $f_0$					&		year, Ref.	\\
\hline
DLCQ					& $5.52$              	&	1988, \cite{ref14:DLCQ}			\\
QSE diagonalization			& $10$                	&	2000, \cite{ref15:QSE}			\\
DMRG					& $9.9816(16)$        	&	2004, \cite{ref16:DMRG}			\\
Monte Carlo cluster			& $10.8^{0.1}_{0.05}$ 	&	2009, \cite{ref6:SchaichLoinaz}		\\
Monte Carlo SLAC derivative		& $10.92(13)$		&	2012, \cite{ref20:SLAC}			\\
Uniform Matrix product states	        & $11.064(20)$        	&	2013, \cite{ref17:Umatrix}		\\
Renormalised Hamiltonian		& $11.88(56)$         	&	2015, \cite{ref18:HT}			\\
Monte Carlo worm			& $11.15(6)(3)$       	&	This work				\\
\end{tabular} 
\end{ruledtabular}
\end{table}

\section{Conclusions}
We decide to quote our final result as:
\begin{equation}
f_0 = 11.15(6)(3).
\label{eq:ffinal}
\end{equation}
We take as central value the mean of \eqref{eq:f1} and \eqref{eq:f2}. The first error is purely statistical, and it is conservatively taken as the biggest one between the two fits. The second error is an estimate of the systematic error  associated with the particular functional form used to fit data.

In Table \ref{tab:review} we summarize some of the latest results for $f_0$ derived with different approaches: the works \cite{ref14:DLCQ,ref15:QSE,ref16:DMRG,ref17:Umatrix,ref18:HT} are based on Hamiltonian truncation (variational) methods, while in \cite{ref20:SLAC} lattice theory is simulated by using non--local SLAC derivative.

We note that our result is compatible with the last four determinations, which come from different methods. We only observe a discrepancy at a $3\sigma$--level with the Monte Carlo results in \cite{ref6:SchaichLoinaz}, where a region of very small $g$--values is reached. For technical reasons, which will be hopefully overcome in the near future, we could not reach this region, but thanks to the worm algorithm our statistical errors are much smaller. We also note that the result of our second fit ($\eta$--parametrization, see Fig. \ref{fig:feta}) has a statistical error comparable with that of \cite{ref17:Umatrix}, and the two results are compatible at $2\sigma$--level. Although we were very conservative in the error estimations, we believe that this work is a step towards a more precise Monte Carlo determination of $f_0$.

Our plans for the next future are to improve this work towards the $g\to 0$ limit with an extended statistics.
\section*{Acknowledgments}
We thank P. Pedroni for suggestions and G. Montagna, B. Pasquini, F. Piccinini and M. Verbeni for a critical reading of the manuscript.
\newpage

\bibliographystyle{apsrev4-1}
\bibliography{bib}

\end{document}